\documentclass{Interspeech}
\usepackage[utf8]{inputenc}
\usepackage{graphicx}
\usepackage{multirow}
\usepackage{tikz}
\usepackage[T1]{fontenc}

\interspeechcameraready

\title{SpeechMLC: Speech Multi-label Classification}

\author[affiliation={1}]{Miseul}{Kim}
\author[affiliation={1}]{Seyun}{Um}
\author[affiliation={1}]{Hyeonjin}{Cha}
\author[affiliation={1}]{Hong-goo}{Kang}

\affiliation{Department of Electrical Engineering}{Yonsei University, Seoul}{South Korea}
\email{\{miseul4345, syum, hcha\}@dsp.yonsei.ac.kr, hgkang@yonsei.ac.kr}
\keywords{speech multi-label classification, human-machine interaction}

\usepackage{comment}
\usepackage{cite}

\begin{document}

\maketitle
\begin{abstract}
\label{0_abstract}
In this paper, we propose a multi-label classification framework to detect multiple speaking styles in a speech sample.
Unlike previous studies that have primarily focused on identifying a single target style, our framework effectively captures various speaker characteristics within a unified structure, making it suitable for generalized human-computer interaction applications.
The proposed framework integrates cross-attention mechanisms within a transformer decoder to extract salient features associated with each target label from the input speech.
To mitigate the data imbalance inherent in multi-label speech datasets, we employ a data augmentation technique based on a speech generation model.
We validate our model's effectiveness through multiple objective evaluations on seen and unseen corpora.
In addition, we provide an analysis of the influence of human perception on classification accuracy by considering the impact of human labeling agreement on model performance.
\end{abstract}

\section{Introduction}
\label{1_intro}

Recent advancements in generative AI models have led to significant improvements in tasks such as expressive text-to-speech (TTS)~\cite{yang2024instructtts, pavankalyan23_interspeech} and text-to-audio (TTA) generation~\cite{liu2024audioldm, kim24e_interspeech}. 
As a result, there is an increasing demand for robust evaluation methods to assess whether synthesized samples accurately capture the intended characteristics.
This has brought increased attention to speaking style classification (SSC), the task of identifying and distinguishing various speaking styles from speech utterances~\cite{veiga2012prosodic}.

A key task in SSC is speech emotion classification~\cite{ranjan24_interspeech, wang23ka_interspeech}, which aims to identify a speaker's emotional state from a given speech sample. 
Research in this area has notably been supported by several publicly available datasets~\cite{busso2008iemocap, 9413391, livingstone2018ryerson}. 
Leveraging their powerful ability to extract localized features from input data, deep learning-based speech emotion classification models have frequently adopted CNN-based frameworks for network design~\cite{lakshmi2023recognition}.
To further enhance intra-class compactness and inter-class separability among features, some methods have utilized contrastive losses during model training~\cite{lian2018speech}.
With the advent of self-supervised learning (SSL) models for speech representations, it has become easier to obtain features that include both acoustic and linguistic information.
Consequently, there has been an increase in approaches for designing speech emotion classification models that incorporate SSL features, replacing traditional statistical acoustic features extracted from waveforms such as mel-spectrograms.

Previous studies have primarily focused on the specific task of emotion detection. 
However, human speech conveys a diverse range of characteristics beyond emotion, highlighting the need for classification models that account for a broader set of attributes. 
Moreover, most existing approaches are limited to single-class classification, limiting their ability to capture the full range of speech characteristics within an utterance.

Driven by growing interest in developing speech generation models that consider diverse speaking factors, several recent works have introduced speech datasets with speaking style annotations.
The LibriTTS-P dataset~\cite{kawamura24_interspeech} provides perceptual annotations of various speaker characteristics for each speaker in the LibriTTS dataset, incorporating a wide range of perceptual labels. 
Additionally, the DreamVoice dataset~\cite{hai24_interspeech} offers a more refined and condensed set of keywords for annotating speaker characteristics.

\begin{figure}[tp]
    \centering
    \includegraphics[width=7cm]{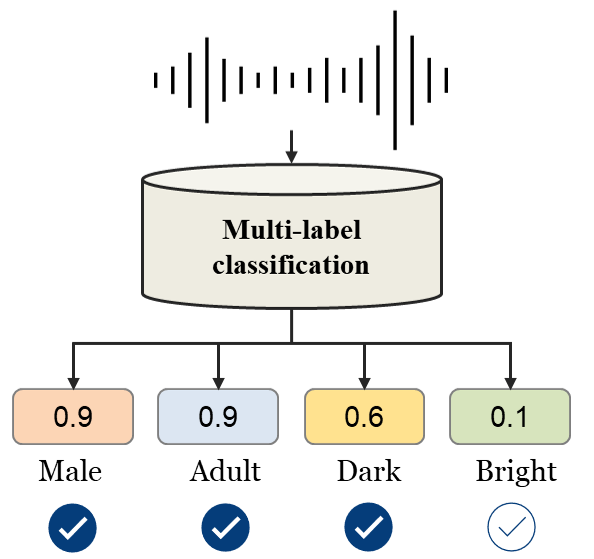}
    \caption{Speech multi-label classification (SpeechMLC). Unlike conventional speech classification tasks, SpeechMLC is designed to predict multiple labels from inputs, capturing various speaker characteristics simultaneously.}
    \label{fig:fig1}
\end{figure}

In this work, we present the first attempt (to the best of our knowledge) to develop a multi-label classification model for speaking styles.  
Our unified framework can identify multiple speaking style classes from a given speech input. 
To address data imbalances, where certain classes are underrepresented in the training data, we employ a data augmentation strategy based on a voice conversion model. 
To assess the generalization ability of our approach, we evaluate the model on both in-domain (seen) and out-of-domain (unseen) corpora, including emotional and cross-lingual data.
In addition, we analyze the impact of human labeling agreement on our model's performance to understand the influence of human perception on classification accuracy. Our key contributions are as follows:
\begin{itemize}
\item Our work introduces the first multi-label classification model for speech input, demonstrating its effectiveness on both in-domain (seen) and out-of-domain (unseen) datasets. 
\item We propose a novel data augmentation approach based on a voice conversion model to enhance the representation of underrepresented classes in the training data. 
\item We investigate the impact of human labeling agreement on classification performance to understand how human perception influences the model's accuracy.
\end{itemize}

\section{Related Works}
\label{2_related}

\begin{figure}[tp]
    \centering
    \includegraphics[width=8cm]{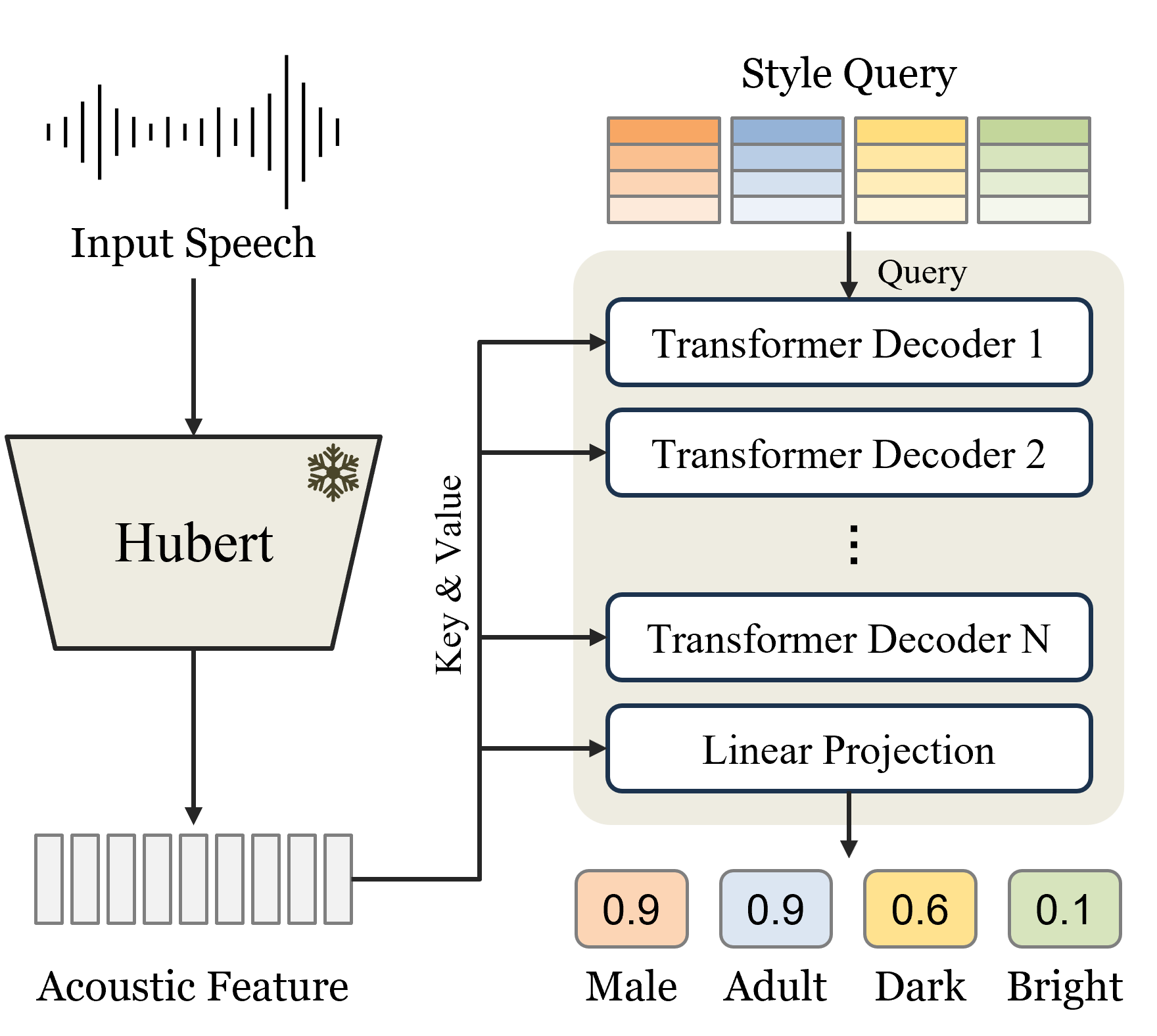}
    \caption{The proposed speech multi-label classification (SpeechMLC) framework. Given an input speech signal, we extract acoustic features using a pre-trained HuBERT backbone, which serve as keys and values in transformer decoder layers. Learnable label embeddings function as the queries (``style queries'').}
    \label{fig:fig2}
\end{figure}

\subsection{Multi-label classification}
Multi-label classification (MLC) is the task of identifying multiple classes present in a given input~\cite{tsoumakas2008multi}.
Unlike single-label classification (SLC), MLC requires simultaneously considering multiple characteristics in a given input sample. 
MLC has been widely explored in the image domain, with applications to object detection, medical image recognition, and scene understanding~\cite{wehrmann2018hierarchical, chai2023multi, zhu2023multi}.
In general, CNN-based models have achieved high accuracy on various datasets.
More advanced models incorporating attention mechanisms have demonstrated further performance improvements~\cite{liu2021query2label}.

Developing speech-based MLC models is challenging compared to the vision-based setting for two main reasons. 
First, while objects in images are clearly distinguishable, the diverse characteristics of speakers are inherently entangled in sequential speech data. 
Second, the development of speech-based MLC models has been hindered by limited availability of publicly annotated datasets, as the labeling process is time-consuming.
However, with the recent release of speech datasets containing samples annotated with multiple descriptive keywords, it has become more feasible to address this task.
Therefore, this work aims to develop an MLC framework specifically tailored to the speech domain.

\subsection{Data augmentation}
A common challenge in multi-label classification is sample imbalance, where many labels are underrepresented in the data compared to a few dominant ones~\cite{tarekegn2021review}.
This imbalance can result in models struggling to learn effective representations for the minority classes.
One way of addressing class imbalance is the use of focal loss, which dynamically adjusts the loss contribution based on the difficulty of each example, thereby mitigating the impact of any data imbalances~\cite{ross2017focal, chen2022mcfl}. 
Another approach is data augmentation, which expands the training dataset by generating additional examples~\cite{song2024toward, ke2019end}. 
Recent studies have utilized pre-trained generation models to expand the amount of available data for model training~\cite{chai2024compositional}. 
In this work, we employ a simplified data augmentation method to enhance the training data, and we assess its effectiveness by evaluating model performance with different amounts of augmented data.
\section{Methodology}
\label{3_method}

Figure~\ref{fig:fig2} illustrates our proposed multi-label classification framework, \textbf{SpeechMLC}.
Motivated by~\cite{liu2021query2label, xie2023category}, we incorporate a cross-attention mechanism in a transformer~\cite{waswani2017attention} decoder to capture representations at the speaking style level.
Given an input speech signal, we first extract acoustic features using a pre-trained HuBERT~\cite{hsu2021hubert} backbone.
These acoustic features are then used as keys and values in all of the transformer decoder layers.
Simultaneously, label embeddings, which we refer to as  \textit{style queries}, serve as the queries in the transformer decoder.
After passing through a final fully-connected (FC) layer, the model outputs logits for each category.
Section~\ref{3.1_speechmlc} provides a detailed description of our framework, while Section~\ref{3.2_data_augmentation} introduces our data augmentation techniques.

\begin{figure}[tp]
    \centering
    \includegraphics[width=8cm]{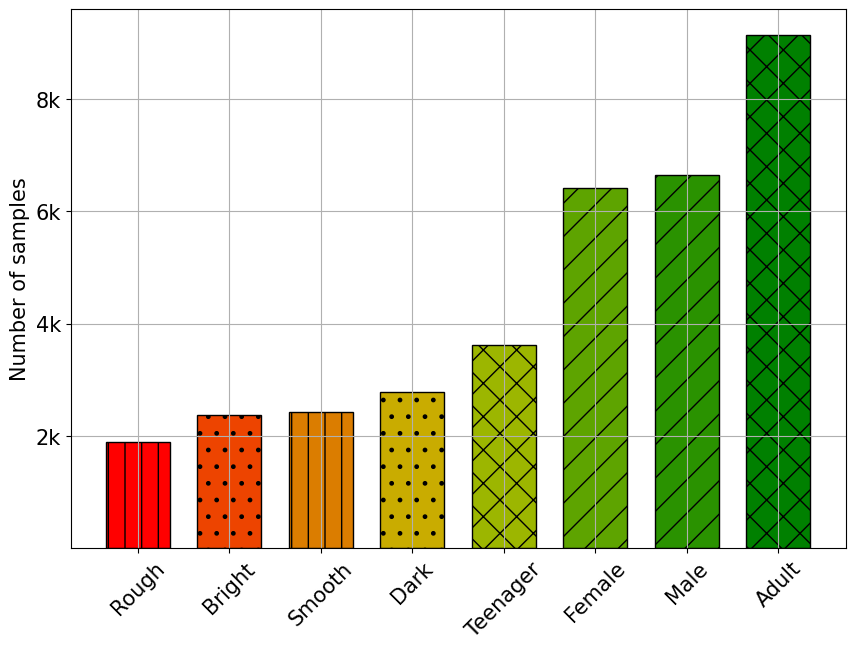}
    \vspace{-20pt}
    \caption{Label distribution of the DreamVoice dataset. It indicates the number of samples per label, highlighting the data imbalance in the training set.} %
    \label{fig:fig3}
\end{figure}

\subsection{SpeechMLC}
\label{3.1_speechmlc}
Our proposed SpeechMLC framework aims to predict the presence of each category in a given speech sample \textbf{x} from a pre-defined set of categories.
Let the total number of categories be $K$ and the corresponding label for $\mathbf{x}$ be $\mathbf{y} =[y_1,...,y_k,...,y_K]$.
For each category $k$, $y_k$ is set to 1 if speech $\mathbf{x}$ exhibits the $k$-th category, and 0 otherwise.
Given input speech $\mathbf{x}$, SpeechMLC predicts the probability of the presence of each category's presence under the supervision provided by the target label set $\mathbf{y}$.
In vision-based multi-label classification, a common approach utilizes a pre-trained single-label image classification model to extract spatial features from the input image~\cite{gupta2024open, sun2022dualcoop}. 
However, there is a lack of publicly available pre-trained speech classification models for extracting analogous features from input speech.
As an alternative, we utilize a pre-trained self-supervised learning (SSL) model for speech representations.
In this study, we adopt the HuBERT model~\cite{hsu2021hubert} for acoustic feature extraction; however, any SSL-based feature extractor can generally be used.

The acoustic features extracted from HuBERT are used as the key and value to the attention mechanisms in a transformer-based decoder.
For the queries, we use learnable label embeddings (\textit{style queries}).
The key, value, and query are then combined using cross-attention mechanisms at each layer of the multi-layer transformer decoder.
During training, each label embedding is updated dynamically, enabling it to learn category-specific features.
As a result, the label embeddings learn to implicitly capture the correlation between the input data and the corresponding labels.
For multi-label classification, each label prediction is treated as a binary classification task. 
Specifically, for each label $k$, its corresponding features are projected to a logit value using a linear projection layer followed by a sigmoid activation function.

\begin{table}[t]
\centering
\renewcommand{\arraystretch}{1.2}
\caption{The ratio of each target label where at least 5 out of 8 annotators reached an agreement in the DreamVoice dataset.}
\vspace{-5pt}
\fontsize{7pt}{8.5pt}\selectfont
\resizebox{\columnwidth}{!}{%
\begin{tabular}{cc|cc}
\hline
Label                        & Ratio (\%) & Label                      & Ratio (\%) \\ \hline
Female                       & 100.00      & Dark                       & 82.26      \\
Male                         & 100.00      & Bright                     & 81.20      \\ \hline
Adult                        & 100.00      & Rough                      & 27.84      \\
\multicolumn{1}{l}{Teenager} & 17.68       & \multicolumn{1}{l}{Smooth} & 44.32       \\ \hline
\end{tabular}%
}
\label{tab1:tab1}
\end{table}

\begin{table*}[t]
\renewcommand{\arraystretch}{1.2}
\caption{Multi-label classification results for ResNet-34 and SpeechMLC in terms of F1-scores.}
\vspace{-5pt}
\begin{tabular}{ccc|cccccccc|c}
\hline
\multicolumn{3}{c|}{Experiments} &
  \multirow{2}{*}{Female} &
  \multirow{2}{*}{Male} &
  \multirow{2}{*}{Adult} &
  \multirow{2}{*}{Teenager} &
  \multirow{2}{*}{Dark} &
  \multirow{2}{*}{Bright} &
  \multirow{2}{*}{Rough} &
  \multirow{2}{*}{Smooth} &
  \multirow{2}{*}{Average} \\ \cline{1-3}
Model                               & Feature    & Augment &        &       &       &       &       &       &       &       &       \\ \hline
ResNet-34                           & Mel80      & x       & 0.901  & 0.917 & 0.402 & 0.402 & 0.418 & 0.435 & 0.388 & 0.461 & 0.541 \\ \hline
\multirow{3}{*}{SpeechMLC}          & Mel80      & x       & 0.855  & 0.950 & 0.706 & 0.709 & 0.863 & 0.767 & 0.460 & 0.528 & 0.730 \\
                                    & HuBERT     & x       & \textbf{0.931}  & \textbf{0.996} & 0.898 & 0.899 & \textbf{0.916} & 0.853 & 0.668 & 0.513 & 0.834 \\
 &
  HuBERT &
  o &
  0.930 &
  0.995 &
  \textbf{0.903} &
  \textbf{0.906} &
  \textbf{0.916} &
  \textbf{0.856} &
  \textbf{0.669} &
  \textbf{0.537} &
  \textbf{0.839} \\ \hline
\end{tabular}%
\label{tab2:tab2}
\end{table*}

\subsection{Data augmentation with voice conversion}
\label{3.2_data_augmentation}
Figure~\ref{fig:fig3} presents the number of samples for each target label in the dataset we use (DreamVoice; see Section~\ref{4.1_dataset}), illustrating the data imbalance that is present.
To mitigate this issue, we applied data augmentation to increase the number of samples for underrepresented labels, particularly focusing on `Rough' and `Smooth'.
To do this, we utilized a voice conversion model; we chose this option rather than text-to-speech (TTS) because TTS models can sometimes be unstable for speech generation (e.g., prone to issues like speech deletion)~\cite{hayashi2020espnet}.
Specifically, we adopted KNN-VC~\cite{baas2023voice}, which is considered state-of-the-art in voice conversion. To ensure sufficient speaker information, we provided the model 60 seconds of target speech as input.
\section{Experimental Setup}
\subsection{Dataset}
\label{4.1_dataset}
We utilized the DreamVoice dataset~\cite{hai24_interspeech} for training, which provides multi-label annotations of speaking styles for 900 speakers from LibriTTS-R and VCTK.
A total of 8 annotators with expertise in speech or audio labeled all samples.
In this work, we focused on four categories from the dataset: gender, age, brightness, and roughness. 
For training simplicity, we merged the `adult' and `senior' categories into a single `adult' category.
For the gender category, samples labeled as `ambiguous' were excluded during training. 
Therefore, our target style categories consist of 8 keywords: [`Female', `Male', `Adult', `Teenager', `Dark', `Bright', `Rough', `Smooth'].

To evaluate our model's performance, we selected a total of 20 unseen speakers from the DreamVoice dataset, leaving 880 speakers for training.
Table~\ref{tab1:tab1} presents the number of samples for each target label where at least 5 out of 8 annotators reached an agreement.
There are significant differences between classes. 
To ensure consistency in measuring classification accuracy across labels during evaluation, we only selected data from the 20 unseen speakers where the number of agreements was greater than 5.

\subsection{Implementation details}
For acoustic feature extraction from input speech, we leveraged a pre-trained HuBERT-base model.\footnote{https://github.com/facebookresearch/fairseq/blob/main/examples/hubert}
We extracted HuBERT features from the 6th layer of the model.
For the transformer decoder, we used 4 decoder layers, each with 8 multi-head attention layers and a dimension of 128.
For style queries, we used queries with a dimension of 128, all with trainable weights. 
To train the model, we used the Adam optimizer~\cite{diederik2014adam} with a learning rate of 0.0001 and a batch size of 64.
For batch feeding, we cropped each sample's duration to 5 seconds; if a sample was shorter than 5 seconds, we padded it with zeros at the end.
For data augmentation, we adopted the kNN-VC model from the official GitHub repository.\footnote{https://github.com/bshall/knn-vc}
We generated a total of 14 hours of samples, by creating source speech and target speaker pairs from the DreamVoice training split.

\section{Experiments}
\subsection{Overall model performance}
Since this work is the first attempt to develop an MLC model using speech signals, no existing baseline models are available for performance comparison.
Therefore, we compared the performance of the proposed model with that of a ResNet-34 architecture~\cite{he2016deep}, which is commonly used in various speech classification tasks.
The ResNet-34 model was also trained on the MLC criterion in the same way as SpeechMLC.
As shown in Table~\ref{tab2:tab2}, SpeechMLC achieves higher F1-scores than ResNet-34 for all classes.
We also observe that using HuBERT features instead of 80-channel mel-spectrograms (Mel80) results in an improved F1-score.
\subsection{Model performance depending on data augmentation}
In Table~\ref{tab2:tab2}, we confirmed the efficacy of data augmentation using F1-score.
Although we trained our model on a dataset augmented with 14 hours of 'Rough' and 'Smooth' utterances, the performance differences were insignificant.
We suspect two reasons for this lack of significant performance improvement. 
1) As shown in Table~\ref{tab1:tab1}, only a small number of voices are classified as 'Rough' or 'Smooth' by at least 5 out of 8 annotators.
2) Voice distortion may have occurred during the voice conversion process, resulting in altered speaking styles for the target speakers.


\subsection{Performance with unseen corpus}
\begin{figure}[tp]
    \centering
    \includegraphics[width=8cm]{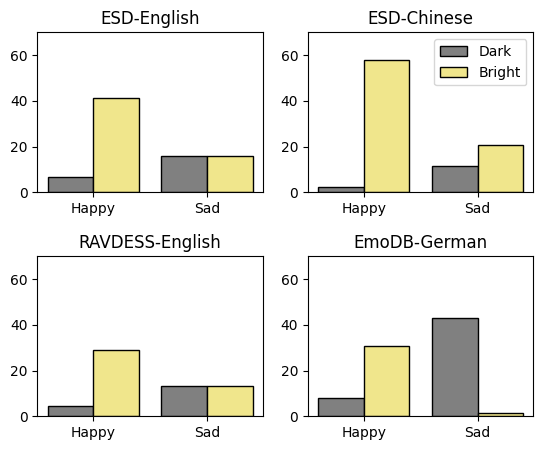}
    \vspace{-20pt}
    \caption{Detection probabilities of SpeechMLC for the presence of ``Dark'' and ``Bright'' on unseen emotional speech datasets in English, Chinese, and German, applied towards utterances annotated as ``Happy'' or ``Sad''.} %
    \label{fig:fig4}
\end{figure}

To evaluate our model's performance in an out-of-domain setting, we looked at the probabilities that it detects the presence of `Dark’ and `Bright’ classes in unseen emotional speech data in Figure~\ref{fig:fig4}.
We performed classification on publicly available emotional speech data in various languages, including English, Chinese, and German.
Specifically, we adopted the ESD dataset~\cite{9413391}, RAVDESS~\cite{livingstone2018ryerson}, and a Berlin EmoDB~\cite{burkhardt2005database}.
Because samples in these datasets do not explicitly contain annotations corresponding to ``Dark'' and ``Bright'', we considered samples with the labels ``Happy'' and ``Sad'', which we judged to have high similarity with ``Dark'' and ``Bright''.
To obtain data with clearly expressed emotions, for RAVDESS, we specifically selected utterances with strong emotional intensity.
For ESD, we report results depending on each language represented in the data (ESD-English, ESD-Chinese).
For all three datasets, we found that the probability of detecting ``Bright'' was higher than that for ``Dark'' in the utterances labeled as ``Happy''.
In EmoDB, the probability of detecting 'Dark' was very high, while 'Brightness' was rarely detected in a sad voice.
On the other hand, we saw similarly low probabilities of detection for both ``Dark'' and ``Bright'' in ``Sad'' voices on ESD-English and RAVDESS.
We suspect that the low probabilities for ``Dark'' in the ``Sad'' utterances are due to many of the voice actors producing speech in a somewhat neutral tone.
In addition, we suspect that the higher probabilities for ``Bright'' compared to ``Dark'' in ESD-Chinese are due to the tonal characteristics of Chinese, which include a wider variety of prosodic features.

\subsection{Impact of human annotator agreement}
\begin{figure}[tp]
    \centering
    \includegraphics[width=8.5cm]{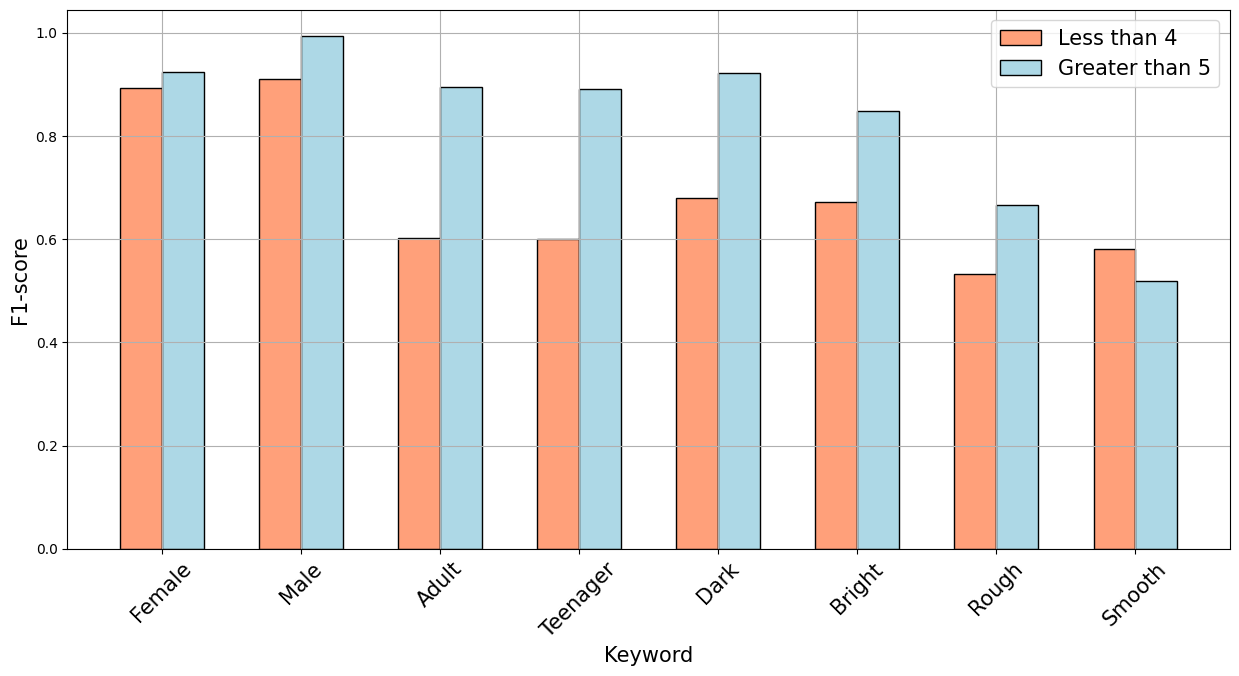}
    \vspace{-20pt}
    \caption{F1-score of SpeechMLC when applied to samples with agreement between 4 or fewer vs. 5 or more annotators.} %
    \label{fig:fig5}
\end{figure}

In speaking style classification datasets, label annotations are based on the subjective decisions of human annotators.
Therefore, the quality of human annotations is crucial in determining the quality of the dataset, and is closely related to model performance.  
To analyze the effect of human annotation on model performance, for all 8 classes, we plot the F1-score of our model when applied to samples with agreement between 4 or fewer vs. 5 or more annotators.
As shown in Figure~\ref{fig:fig5}, we observed that lower annotator agreement in labeling corresponds to lower model performance, implying that using datasets annotated with low-quality labels can make it difficult to evaluate model performance accurately.
This emphasizes the need for creating more high-quality speech datasets with speaking style annotations in order to develop reliable models that can capture these diverse characteristics.

\section{Conclusion}
In this paper, we introduced SpeechMLC, a multi-label classification framework for detecting multiple speaking styles in a speech utterance.
To the best of our knowledge, this is the first work that attempts to solve this task.
Our model combines HuBERT features with learnable style queries using cross-attention in a transformer decoder-based architecture, and outperforms a ResNet-34 baseline on the multi-label classification task.
We also addressed class imbalance problems in our dataset by generating more samples for underrepresented classes with a voice conversion model, and studied our model's generalizability to unseen emotional speech datasets across 3 different languages.
Finally, we analyzed how our model's performance is affected by the level of human annotator agreement in the data, highlighting the need for further work and data in this domain in order to develop machine learning models that can reliably capture diverse stylistic characteristics of speech.
\clearpage

\bibliographystyle{IEEEtran}
\bibliography{mybib}

\begin{thebibliography}{10}
\providecommand{\url}[1]{#1}
\csname url@samestyle\endcsname
\providecommand{\newblock}{\relax}
\providecommand{\bibinfo}[2]{#2}
\providecommand{\BIBentrySTDinterwordspacing}{\spaceskip=0pt\relax}
\providecommand{\BIBentryALTinterwordstretchfactor}{4}
\providecommand{\BIBentryALTinterwordspacing}{\spaceskip=\fontdimen2\font plus
\BIBentryALTinterwordstretchfactor\fontdimen3\font minus \fontdimen4\font\relax}
\providecommand{\BIBforeignlanguage}[2]{{%
\expandafter\ifx\csname l@#1\endcsname\relax
\typeout{** WARNING: IEEEtran.bst: No hyphenation pattern has been}%
\typeout{** loaded for the language `#1'. Using the pattern for}%
\typeout{** the default language instead.}%
\else
\language=\csname l@#1\endcsname
\fi
#2}}
\providecommand{\BIBdecl}{\relax}
\BIBdecl

\bibitem{yang2024instructtts}
D.~Yang, S.~Liu, R.~Huang, C.~Weng, and H.~Meng, ``Instructtts: Modelling expressive tts in discrete latent space with natural language style prompt,'' \emph{IEEE/ACM Transactions on Audio, Speech, and Language Processing}, 2024.

\bibitem{pavankalyan23_interspeech}
T.~{Pavan Kalyan}, P.~Rao, P.~Jyothi, and P.~Bhattacharyya, ``Narrator or character: Voice modulation in an expressive multi-speaker tts,'' in \emph{Interspeech 2023}, 2023, pp. 4808--4812.

\bibitem{liu2024audioldm}
H.~Liu, Y.~Yuan, X.~Liu, X.~Mei, Q.~Kong, Q.~Tian, Y.~Wang, W.~Wang, Y.~Wang, and M.~D. Plumbley, ``Audioldm 2: Learning holistic audio generation with self-supervised pretraining,'' \emph{IEEE/ACM Transactions on Audio, Speech, and Language Processing}, 2024.

\bibitem{kim24e_interspeech}
M.~Kim, S.-W. Chung, Y.~Ji, H.-G. Kang, and M.-S. Choi, ``Speak in the scene: Diffusion-based acoustic scene transfer toward immersive speech generation,'' in \emph{Interspeech 2024}, 2024, pp. 4883--4887.

\bibitem{veiga2012prosodic}
A.~Veiga, D.~Celorico, J.~Proen{\c{c}}a, S.~Candeias, and F.~Perdig{\~a}o, ``Prosodic and phonetic features for speaking styles classification and detection,'' in \emph{Advances in Speech and Language Technologies for Iberian Languages: IberSPEECH 2012 Conference, Madrid, Spain, November 21-23, 2012. Proceedings}.\hskip 1em plus 0.5em minus 0.4em\relax Springer, 2012, pp. 89--98.

\bibitem{ranjan24_interspeech}
S.~Ranjan, R.~Chakraborty, and S.~K. Kopparapu, ``Reinforcement learning based data augmentation for noise robust speech emotion recognition,'' in \emph{Interspeech 2024}, 2024, pp. 1040--1044.

\bibitem{wang23ka_interspeech}
S.~Wang, J.~Guðnason, and D.~Borth, ``Learning emotional representations from imbalanced speech data for speech emotion recognition and emotional text-to-speech,'' in \emph{Interspeech 2023}, 2023, pp. 351--355.

\bibitem{busso2008iemocap}
C.~Busso, M.~Bulut, C.-C. Lee, A.~Kazemzadeh, E.~Mower, S.~Kim, J.~N. Chang, S.~Lee, and S.~S. Narayanan, ``Iemocap: Interactive emotional dyadic motion capture database,'' \emph{Language resources and evaluation}, vol.~42, pp. 335--359, 2008.

\bibitem{9413391}
K.~Zhou, B.~Sisman, R.~Liu, and H.~Li, ``Seen and unseen emotional style transfer for voice conversion with a new emotional speech dataset,'' in \emph{ICASSP 2021 - 2021 IEEE International Conference on Acoustics, Speech and Signal Processing (ICASSP)}, 2021, pp. 920--924.

\bibitem{livingstone2018ryerson}
S.~R. Livingstone and F.~A. Russo, ``The ryerson audio-visual database of emotional speech and song (ravdess): A dynamic, multimodal set of facial and vocal expressions in north american english,'' \emph{PloS one}, vol.~13, no.~5, p. e0196391, 2018.

\bibitem{lakshmi2023recognition}
K.~L. Lakshmi, P.~Muthulakshmi, A.~A. Nithya, R.~B. Jeyavathana, R.~Usharani, N.~S. Das, and G.~N.~R. Devi, ``Recognition of emotions in speech using deep cnn and resnet,'' \emph{Soft Computing}, pp. 1--17, 2023.

\bibitem{lian2018speech}
Z.~Lian, Y.~Li, J.~Tao, and J.~Huang, ``Speech emotion recognition via contrastive loss under siamese networks,'' in \emph{Proceedings of the Joint Workshop of the 4th Workshop on Affective Social Multimedia Computing and First Multi-Modal Affective Computing of Large-Scale Multimedia Data}, 2018, pp. 21--26.

\bibitem{kawamura24_interspeech}
M.~Kawamura, R.~Yamamoto, Y.~Shirahata, T.~Hasumi, and K.~Tachibana, ``Libritts-p: A corpus with speaking style and speaker identity prompts for text-to-speech and style captioning,'' in \emph{Interspeech 2024}, 2024, pp. 1850--1854.

\bibitem{hai24_interspeech}
J.~Hai, K.~Thakkar, H.~Wang, Z.~Qin, and M.~Elhilali, ``Dreamvoice: Text-guided voice conversion,'' in \emph{Interspeech 2024}, 2024, pp. 4373--4377.

\bibitem{tsoumakas2008multi}
G.~Tsoumakas and I.~Katakis, ``Multi-label classification: An overview,'' \emph{Data Warehousing and Mining: Concepts, Methodologies, Tools, and Applications}, pp. 64--74, 2008.

\bibitem{wehrmann2018hierarchical}
J.~Wehrmann, R.~Cerri, and R.~Barros, ``Hierarchical multi-label classification networks,'' in \emph{International conference on machine learning}.\hskip 1em plus 0.5em minus 0.4em\relax PMLR, 2018, pp. 5075--5084.

\bibitem{chai2023multi}
Y.~Chai, H.~Liu, J.~Xu, S.~Samtani, Y.~Jiang, and H.~Liu, ``A multi-label classification with an adversarial-based denoising autoencoder for medical image annotation,'' \emph{ACM Transactions on Management Information Systems}, vol.~14, no.~2, pp. 1--21, 2023.

\bibitem{zhu2023multi}
K.~Zhu, M.~Fu, and J.~Wu, ``Multi-label self-supervised learning with scene images,'' in \emph{Proceedings of the IEEE/CVF International Conference on Computer Vision}, 2023, pp. 6694--6703.

\bibitem{liu2021query2label}
S.~Liu, L.~Zhang, X.~Yang, H.~Su, and J.~Zhu, ``Query2label: A simple transformer way to multi-label classification,'' \emph{arXiv preprint arXiv:2107.10834}, 2021.

\bibitem{tarekegn2021review}
A.~N. Tarekegn, M.~Giacobini, and K.~Michalak, ``A review of methods for imbalanced multi-label classification,'' \emph{Pattern Recognition}, vol. 118, p. 107965, 2021.

\bibitem{ross2017focal}
T.-Y. Ross and G.~Doll{\'a}r, ``Focal loss for dense object detection,'' in \emph{proceedings of the IEEE conference on computer vision and pattern recognition}, 2017, pp. 2980--2988.

\bibitem{chen2022mcfl}
L.~Chen, J.~Song, X.~Zhang, and M.~Shang, ``Mcfl: multi-label contrastive focal loss for deep imbalanced pedestrian attribute recognition,'' \emph{Neural Computing and Applications}, vol.~34, no.~19, pp. 16\,701--16\,715, 2022.

\bibitem{song2024toward}
H.~Song, M.~Kim, and J.-G. Lee, ``Toward robustness in multi-label classification: A data augmentation strategy against imbalance and noise,'' in \emph{Proceedings of the AAAI Conference on Artificial Intelligence}, vol.~38, no.~19, 2024, pp. 21\,592--21\,601.

\bibitem{ke2019end}
X.~Ke, J.~Zou, and Y.~Niu, ``End-to-end automatic image annotation based on deep cnn and multi-label data augmentation,'' \emph{IEEE Transactions on Multimedia}, vol.~21, no.~8, pp. 2093--2106, 2019.

\bibitem{chai2024compositional}
Y.~Chai, Z.~Li, J.~Liu, L.~Chen, F.~Li, D.~Ji, and C.~Teng, ``Compositional generalization for multi-label text classification: A data-augmentation approach,'' in \emph{Proceedings of the AAAI Conference on Artificial Intelligence}, vol.~38, no.~16, 2024, pp. 17\,727--17\,735.

\bibitem{xie2023category}
C.~Xie, F.~Zeng, Y.~Hu, S.~Liang, and Y.~Wei, ``Category query learning for human-object interaction classification,'' in \emph{Proceedings of the IEEE/CVF Conference on Computer Vision and Pattern Recognition}, 2023, pp. 15\,275--15\,284.

\bibitem{waswani2017attention}
A.~Waswani, N.~Shazeer, N.~Parmar, J.~Uszkoreit, L.~Jones, A.~Gomez, L.~Kaiser, and I.~Polosukhin, ``Attention is all you need,'' in \emph{NIPS}, 2017.

\bibitem{hsu2021hubert}
W.-N. Hsu, B.~Bolte, Y.-H.~H. Tsai, K.~Lakhotia, R.~Salakhutdinov, and A.~Mohamed, ``Hubert: Self-supervised speech representation learning by masked prediction of hidden units,'' \emph{IEEE/ACM transactions on audio, speech, and language processing}, vol.~29, pp. 3451--3460, 2021.

\bibitem{gupta2024open}
R.~Gupta, M.~N. Rizve, J.~Unnikrishnan, A.~Tawari, S.~Tran, M.~Shah, B.~Yao, and T.~Chilimbi, ``Open vocabulary multi-label video classification,'' in \emph{European Conference on Computer Vision}.\hskip 1em plus 0.5em minus 0.4em\relax Springer, 2024, pp. 276--293.

\bibitem{sun2022dualcoop}
X.~Sun, P.~Hu, and K.~Saenko, ``Dualcoop: Fast adaptation to multi-label recognition with limited annotations,'' \emph{Advances in Neural Information Processing Systems}, vol.~35, pp. 30\,569--30\,582, 2022.

\bibitem{hayashi2020espnet}
T.~Hayashi, R.~Yamamoto, K.~Inoue, T.~Yoshimura, S.~Watanabe, T.~Toda, K.~Takeda, Y.~Zhang, and X.~Tan, ``Espnet-tts: Unified, reproducible, and integratable open source end-to-end text-to-speech toolkit,'' in \emph{ICASSP 2020-2020 IEEE international conference on acoustics, speech and signal processing (ICASSP)}.\hskip 1em plus 0.5em minus 0.4em\relax IEEE, 2020, pp. 7654--7658.

\bibitem{baas2023voice}
M.~Baas, B.~van Niekerk, and H.~Kamper, ``Voice conversion with just nearest neighbors,'' \emph{arXiv preprint arXiv:2305.18975}, 2023.

\bibitem{diederik2014adam}
P.~K. Diederik, ``Adam: A method for stochastic optimization,'' \emph{(No Title)}, 2014.

\bibitem{he2016deep}
K.~He, X.~Zhang, S.~Ren, and J.~Sun, ``Deep residual learning for image recognition,'' in \emph{Proceedings of the IEEE conference on computer vision and pattern recognition}, 2016, pp. 770--778.

\bibitem{burkhardt2005database}
F.~Burkhardt, A.~Paeschke, M.~Rolfes, W.~F. Sendlmeier, B.~Weiss \emph{et~al.}, ``A database of german emotional speech.'' in \emph{Interspeech}, vol.~5, 2005, pp. 1517--1520.

\end{thebibliography}
\end{document}